\documentclass[pre,twocolumn,floatfix]{revtex4} 

\usepackage{epsfig}
\usepackage{graphicx}

\begin{document}


\author{B. Utter}
\email{utter@phy.duke.edu}
\altaffiliation{Present address: Dept. of Physics, Box 90305, 
Duke University, Durham, NC 27708}
\author{E. Bodenschatz}
\email{eb22@cornell.edu}
\affiliation{Laboratory of Atomic and Solid State Physics, Cornell University,
	Ithaca, NY  14853}

\title{Dynamics of Low Anisotropy Morphologies in Directional Solidification}
\date{\today}

\renewcommand{\textfraction}{0.15}
\renewcommand{\topfraction}{0.95}
\renewcommand{\bottomfraction}{0.95}
\setcounter{bottomnumber}{2} 
\setcounter{topnumber}{2}
\renewcommand{\floatpagefraction}{0.75}

\newcommand{\ea}{\textit{et al.}}

\begin{abstract}       
We report experimental results on quasi-two-dimensional
diffusion limited growth in directionally solidified
succinonitrile with small amounts of poly(ethylene oxide),
acetone, or camphor as a solute. Seaweed growth, or dense branching
morphology, is
selected by growing grains close to the $\{$111$\}$ plane, where
the in-plane surface tension is nearly isotropic. The  observed growth
morphologies are very sensitive to small 
anisotropies in surface tension caused by misorientations from the $\{$111$\}$ plane.
 Different seaweed morphologies are
found, including the degenerate, the stabilized, and the strongly
tilted seaweeds.  The degenerate seaweeds show a limited fractal scaling
range and, with increased undercooling, suggests a transition 
from ``fractal'' to ``compact'' seaweed. 
Strongly tilted seaweeds demonstrate a
significant twofold anisotropy.
In addition,  seaweed-dendrite transitions are observed in 
low anisotropy growth. 
\end{abstract}         


\maketitle





\section{INTRODUCTION}

It is well known that surface tension anisotropy plays a crucial
role in the formation of cells and dendrites in solidification
microstructures \cite{reviews-PRE-01}. Early on, for isotropic growth, theory
found \cite{Ivantsov:47:Temperature} that  the 
speed and tip radius of cellular growth were nonunique, while
experiment showed clear selection \cite{Glicksman.ea:76:Dendritic}. 
The break-through to this
puzzle came when it was shown that a small amount of anisotropy
acts as a singular perturbation destroying the non-uniqueness of
the selected tip
\cite{reviews-PRE-01}.

Cells and dendrites have been studied extensively, but the study
of nearly isotropic growth in solidification has received less
attention. Without anisotropy the growth is characterized
by frequent random tip splitting leading to a disordered pattern. This
situation has been coined seaweed growth \cite{ihle-mk-93-4} or
dense branching morphology (DBM) \cite{Ben-Jacob.ea:86:Formation}.
Similar patterns are observed in other growth systems which lack
anisotropy, most notably viscous fingering (Hele-Shaw flow) 
\cite{McCloud.ea:95:Experimental,Ben-Jacob.ea:90:formation}, but
also in such different systems as growth of bacterial
colonies~\cite{Matsuyama.ea:93:Fractal,wakita-1998},
electrodeposition~\cite{Ben-Jacob.ea:90:formation,Zik.ea:96:Electrodeposition},
annealing of magnetic films~\cite{Shang:96:Pattern-formation}, and
drying water films~\cite{Samid-Merzel.ea:98:Pattern}. In fact, in
viscous fingering experiments, it was found that introducing
anisotropy can stabilize the tips and induce dendrites
\cite{Ben-Jacob.ea:85:Experimental}.

In this article we report experimental results on weakly
anisotropic growth in directionally solidified succinonitrile (SCN) with
small amounts of poly(ethylene oxide), acetone, or camphor as a
solute. As described in Section~\ref{sec-expt}, the quasi-2D sample is
oriented close to the \{111\} plane leading to a
nearly isotropic surface tension. Weak deviations from the
$\{111\}$ orientation are found to introduce anisotropies and 
profoundly affect the
tip dynamics of the solidification front. 
These deviations are expected for experimental solidification studies 
using model alloys since precise control of sample orientation is not 
currently possible. 
Different types of seaweeds are observed,
depending on the weak
anisotropy: degenerate seaweeds that can lead to 
alternating tip splitting \cite{Utter.ea:01:Alternating},
stabilized seaweeds, and strongly tilted seaweeds which reveal
a large twofold anisotropy.

In addition, we explore the existence of fractal growth in
degenerate seaweeds at low speeds and find that seaweeds in directional
solidification do not appear to be fractal over a significant
range of length scales. We also report results on transitions
between seaweed and dendrite growth. 

Anisotropy in solidification originates from the capillary length
which is proportional to the surface stiffness
\begin{equation}
\label{sstiff-formula}
\tilde{\gamma}(\hat{\bf n}) = \gamma(\hat{\bf n}) +
\frac{\partial^2 \gamma(\hat{\bf n})}{\partial \alpha^2}
\end{equation}
where $\gamma$
is the surface tension and
$\alpha$ is the angle between the normal to the interface
$\hat{\bf n}$ and the pulling
direction~\cite{Akamatsu.ea:95:Symmetry-broken}.

The origin of the surface tension anisotropy is the underlying crystalline
structure of the growing solid.  Growth is preferred along the crystalline
axes and, in two dimensions, a seed grain will typically grow outward as a
four-armed ``snowflake''.
In directional solidification, in which growth
is forced along a particular direction, the arms or dendrites are tilted
in a direction between the crystalline axis and the imposed temperature
gradient.

The effective in-plane anisotropy depends not only on the crystal
itself, but also on the orientation of the crystal with respect to
the growth direction.  When grown in the $\{111\}$ plane,
the surface tension is nearly isotropic, leading to seaweed growth or DBM
\cite{Akamatsu.ea:95:Symmetry-broken}.

Mathematically, the surface tension can be represented in 3D as
\begin{equation}
\label{stens-formula}
\gamma ( \hat{\bf n} ) = \gamma_0 \left[ 1 +
\epsilon_0 ( n_1^4 + n_2^4 + n_3^4 ) \right]
\end{equation}
where $\gamma_0$ is the isotropic part of the surface tension and
$\epsilon_0$ is the degree of anisotropy\cite{Akamatsu.ea:95:Symmetry-broken}.
The anisotropy has been measured as $\epsilon_0 = 0.0055$ in
SCN\cite{Muschol.ea:92:Surface}.
$n_1, n_2,$ and $n_3$ are the
components of a unit vector $\hat{\bf n}$ that parametrizes the function
in three dimensions.
$\gamma ( \hat{\bf n} )$ is the magnitude of the surface tension for a
surface oriented so that its normal is along $\hat{\bf n}$.
This approximation of the actual surface tension
looks somewhat like a rounded cube in three dimensions
for succinonitrile and has the expected 
sixfold symmetry for a cubic crystal.

In directional solidification, the sample is constrained to grow
within a particular plane, so the possible growth surfaces have orientations
$\hat{\bf n}$ perpendicular to the
interface and lying in the plane of the sample.
Constraining $\hat{\bf n}$ to lie in
a plane is equivalent to taking a particular slice through this
3D surface tension plot.
Changing the orientation of the crystal changes
the shape and magnitude of $\gamma (\hat{\bf n})$ and
$\tilde{\gamma}(\hat{\bf n})$
in the sample plane \cite{Akamatsu.ea:98:Anisotropy-driven}.

\begin{figure}
\includegraphics[width=3.3in]{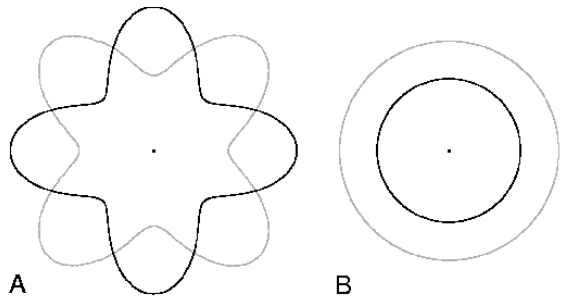}
\caption{\label{stens100-111} 
Using equations~\ref{sstiff-formula} and~\ref{stens-formula},
the surface stiffness
(grey, $\gamma_0$=2, $\epsilon_0$=0.1) and the anisotropic part of the
surface tension
(black, $\gamma ( \hat{\bf n} ) - \gamma_0$, with
$\gamma_0 = 1$ and $\epsilon_0$=2.75) are shown for the (A) \{100\}
and (B) \{111\} planes.
There is significant fourfold anisotropy in the \{100\} plane 
while growth in the \{111\} plane is
isotropic.  Note, the parameters $\gamma_0$ and $\epsilon_0$ are chosen
to emphasize the anisotropy in the surface tension.
}
\end{figure}

Figure~\ref{stens100-111} shows examples of these 2D slices in
the \{100\} plane and the \{111\} plane.
In these cases, the surface stiffness (grey)
has the same symmetry as the surface tension (black).
They are $90^\circ$ out of phase and fingers tend to grow
towards maximum surface tension.
If a crystal in this
orientation was forced to grow upwards, Fig.~\ref{stens100-111}A would
produce stable dendrites with sidebranches at approximately right angles.
We could also rotate the sample (and hence the surface tension plot)
in the plane to produce tilted dendrites.
Without anisotropy of surface tension, Fig.~\ref{stens100-111}B,
the tip is unstable and the growth lacks the apparent
orientation observed in traditional growth morphologies.

\begin{figure}
\includegraphics[width=3.4in]{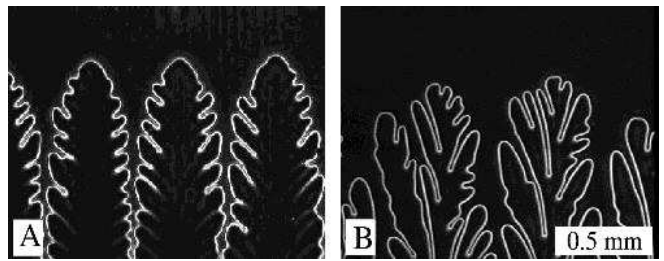}%
\caption{\label{seaweed} 
(A) A dendrite and (B) seaweed structure which differ only in
crystalline orientation.   The white line indicates the
solid-liquid interface.  The solid grows upwards into the undercooled
melt.  The thermal gradient (18 K/cm), concentration (0.25\% SCN-PEO),
and growth
velocity (2.71 $\mu$m/s) are identical in both pictures.
}
\end{figure}

Figure~\ref{seaweed}A and B show experimental pictures of solids
oriented approximately as shown in
Fig.~\ref{stens100-111}A and B respectively.
Seaweed structures (\ref{seaweed}B) are very
disordered compared to more familiar arrays of cells and dendrites
(\ref{seaweed}A).
Note that Fig.~\ref{seaweed}A and B show different seed crystals of 
the same sample
grown under the same growth conditions, illustrating that
it is the crystalline orientation that causes the observed difference.

Although there have been a couple thorough experimental
investigations of the seaweed morphology in directional
solidification\cite{Akamatsu.ea:95:Symmetry-broken,Akamatsu.ea:98:Anisotropy-driven},
very little work has been done on the tip dynamics and the effect of the small
misorientations from the $\{111\}$ plane that are present in any experimental
study.

Previous experiments and simulations
on the seaweed morphology
have focused on the magnitude of the anisotropy, stability of dendrites,
and the orientation of anisotropic crystals.
In particular, Akamatsu and Faivre have performed directional solidification
experiments studying  the effect of surface tension anisotropy and grain
orientation on
morphology\cite{Akamatsu.ea:95:Symmetry-broken,Akamatsu.ea:98:Anisotropy-driven}.
Ihle and
M$\ddot{\mathrm{u}}$ller-Krumbhaar have used numerical simulations
to study seaweeds\cite{ihle-mk-93-4}, including the seaweed-dendrite transition
with increasing anisotropy.
Attempts to vary the anisotropy in simulations \cite{ihle-mk-93-4}
and experiments \cite{Honjo.ea:86:Irregular,Honjo.ea:87:Phase} showed that tip
splitting growth was found when noise was increased.

Brener \ea~propose a morphology diagram involving
the degree of anisotropy and the undercooling \cite{brener-96-8}.
In this diagram,
they distinguish between seaweed structures at low anisotropy and
dendritic structures at high anisotropy and between fractal growth at
low undercooling and compact growth at large undercooling.
They theorize that the fractal structure forms because
tip splitting occurs randomly when the strength of the thermal noise
is large enough to destabilize the
tip \cite{brener-96-8,Brener.ea:94:Fluctuation}.

Honjo \ea~claimed
the first DLA-like crystal growth using NH$_4$Cl crystals radially grown from
solution and found a fractal dimension $D_f$ = 1.671
with about 1 order of magnitude in
length scales \cite{Honjo.ea:86:Irregular}.
Ihle and
M$\ddot{\mathrm{u}}$ller-Krumbhaar have used numerical simulations
to study seaweed morphology
and find $D_f$ = 1.70 $\pm$ 0.03 \cite{ihle-mk-93-4}.
M$\ddot{\mathrm{u}}$ller-Krumbhaar \ea~reconfirmed these results,
$1.66 \leq D_f \leq 1.73$, for a seaweed growth at low
undercooling \cite{Muller-Krumbhaar.ea:96:Morphology}.
Honjo \ea's results
were performed for seaweeds at a particular undercooling
and therefore do not test Brener's predictions of a transition
to compact growth with increased undercooling.  
Ihle and M$\ddot{\mathrm{u}}$ller-Krumbhaar used
three undercoolings and found the
fractal dimension to be approximately constant.  Their scaling range is
not more than one decade and simulations are performed at zero
imposed anisotropy which we are not able to obtain experimentally.
Simulations by Sasikimar and Sreenivasen show an increase in fractal
dimension from 1.6 to 2 with increased undercooling \cite{Sasikumar.ea:94:Two}.
Our results suggest a transition from
fractal to compact growth, but we find that there is not a
significant range of fractal scaling.

At higher anisotropies, the noise is no longer able to destabilize the tip,
but might still be important in inducing sidebranching.  Dynamic
studies of the seaweed morphology might offer more information about the
role of noise in solidification.

No systematic study
has been concerned with  the dynamics of the tip splitting events or
the effect of misorientations from the \{111\} plane.  This seems particularly
important in dense branching morphology 
as slight  misorientations lead to finite anisotropies
to the nominally isotropic case.
In contrast, slight variations on an anisotropic growth such as that in
Fig.~\ref{stens100-111}A would likely  be weak.
We discuss the implications of these misorientations below.

Low anisotropy systems can be very instructive in understanding  the 
transition from seaweeds to dendrites.  This might be particularly 
important for situations where  competing 
anisotropies  nearly balance, such as cases where the kinetic
anisotropy favors 
a different direction than the surface tension anisotropy
\cite{Liu.ea:90:Generic,Ihle:00:Competition}.

This paper is organized as follows:  In Section~\ref{sec-expt} we describe
the experimental apparatus and techniques. In Section \ref{sec-seaweeds},
we characterize three different types of seaweed growth which result from
small anisotropies.  In \ref{sec-fractal}, we study the fractal dimension
of the degenerate seaweed.  In Section~\ref{sec-transitions} we
study seaweed-dendrite transitions for low anisotropy growth.  We conclude in
Section~\ref{sec-conclusions}.

\section{EXPERIMENTAL METHODS}

\label{sec-expt}

\begin{figure}
\includegraphics[width=3.1in]{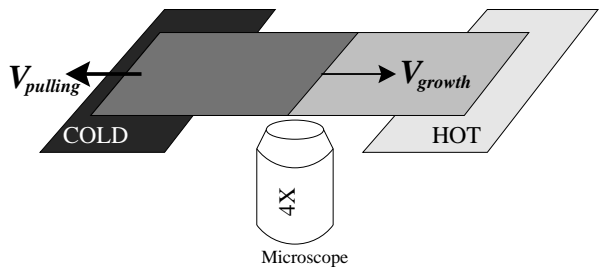}%
\caption{\label{experiment} 
Directional solidification schematic.  A quasi-2D
sample is pulled through a linear
temperature gradient.  The growing
interface is stationary in the lab frame
and is observed through a microscope.
}
\end{figure}

The experiment is performed with a  traditional directional
solidification apparatus \cite{jackson-and-hunt}
in which a thin sample $(13~cm \times 1.5~cm \times (5-60)~{\mu}m)$ is pulled
through a linear temperature gradient at a constant pulling velocity
as shown in Fig.~\ref{experiment}.
After an initial transient, the average speed of the solidification
front is equal to the
pulling speed, set by
a linear stepping motor with 4 nm step size.

The cell consists of two glass plates glued together and filled
with the sample.  The glass plates are cleaned in stages using detergent,
acetone, methanol, an acid solution (sulfuric acid and NoChromix 
(Godax Laboratories, Inc.)),
and distilled water.  The glue used is the epoxy 
Torr-Seal (Varian Vacuum Products).
The nominal
cell depth is set by a Mylar (DuPont)
spacer which can be obtained in a wide
range of thicknesses with good uniformity.

In each set of runs, the temperature gradient is maintained at
a fixed value between 3 and 50 K/cm with a
stability of $\pm 2$ mK possible on each side.   The temperatures of the
hot and cold sides are above and below the equilibrium melting
temperature of $\approx 58^\circ$
so that the solid-liquid interface remains within the gap between the
temperature controlled blocks.
In the most recent design, circular samples are used to allow the cell 
to be rotated within
the sample plane between runs.  This allows for some control
over sample orientation.

The sample used is an alloy of succinonitrile (SCN) and a small amount of
added solute.  The solutes used in the present study are
either 0.25\% poly(ethylene oxide) (PEO)\cite{peo-info},
1.5\% acetone (ACE), or 1.3\% camphor (CAM).
The diffusivities D and partition coefficients k are listed in
Table~\ref{alloy-properties} with the solute concentrations C and
sample thicknesses d used for these results.
The SCN is
purified by sublimation and the samples are mixed, degassed, and vacuum filled
under inert atmosphere to avoid possible contamination.
The melting temperature of the purified material is $58.05 \pm 0.03 ^\circ C$
which corresponds to a purity of 99.98\% \cite{purity}.
Further details
on sample preparation and cell construction will be presented elsewhere
\cite{in-prep}.

\begin{table}[tb]
\begin{ruledtabular}
\begin{tabular}{|l|c|c|c|}
            & SCN-ACE   & SCN-CAM   & SCN-PEO  \\
D ($\mu$m$^2$/s)    & 1270$^a$  & 300$^b$   & 80    \\
k           & 0.1$^a$   & 0.33$^c$  & 0.01  \\
C (weight \%)       & 1.5\%     & 1.3\%     & 0.25\% \\
d ($\mu$m)      & 20        & 22        & 60 \\
\end{tabular}
\caption{
Properties of samples used in this study.
Succinonitrile (SCN) alloys with acetone (ACE), camphor (CAM) and 
poly(ethylene oxide) (PEO) 
as solutes.  Diffusivity D and partition coefficient k are given.
Solute concentration C and sample thickness d used in these
results are also listed.
$a$) Reference \protect \cite{Chopra.ea:88:Measurement}.
$b$) Reference \protect \cite{Sato.ea:87:Experiments}.
$c$) Reference \protect \cite{Taenaka.ea:89:Equilibrium}.
}
\end{ruledtabular}
\label{alloy-properties}

\end{table}

The liquid-solid interface is observed with phase contrast
or Hoffman modulation contrast microscopy.  Sequences
of images are recorded using a CCD camera with a framegrabber or time
lapse video.  Particularly with phase contrast images, such as those in
Fig.~\ref{seaweed}, the interface can then be
easily extracted for further analysis.

To initiate growth, the sample is melted completely and
quenched, seeding a number of grains.
One grain with the desired orientation is selected and all others
are melted off so
the chosen grain can grow and fill the width of the cell.
This is most easily accomplished in SCN-PEO samples since the
attached dye group on the poly(ethylene oxide) \cite{peo-info}
allows us to melt off undesirable grains by locally heating
with an argon laser beam.
The selected grain can then be maintained so runs of different growth speeds
can be performed at the same crystalline orientation.

It is important to
start with a single grain since
dendrites grow at lower undercooling and
typically overtake seaweeds during solidification.
It is common after a run
to have a few subgrains indicating that neigboring lobes can shift slightly
with respect to each other\cite{Akamatsu.ea:95:Symmetry-broken}.
We don't observe any variation in growth morphology after the initial
transient due to the formation of these subgrains.

Before each run, the sample is kept stationary ($V = 0$)
for a sufficient time to
equilibrate the impurity concentration in the liquid and 
create a flat interface.  This allows accurate
measurement of the initial instability wavelength of the flat interface
$\lambda_f$ which results from the Mullins-Sekerka instability
\cite{Mullins.ea:64:Stability}.

Finding an appropriate grain is an experimental
challenge, as the random seeding process gives only a 1/1600
chance of orienting the grain within $1^\circ$ of the \{111\} plane
\cite{random-seeding}.  It
has already been noted that seaweeds exist only within $5^\circ$ of the \{111\}
orientation\cite{Akamatsu.ea:95:Symmetry-broken}.
This $5^\circ$ limit likely depends on the alloy and
concentration used, which appear to affect the degree of anisotropy
in our observations.
However, assuming that limit of seaweed stability,
there is a probability of 1/66 to seed seaweed
growth but only 1 in 25 seaweeds will be within $1^\circ$ of the \{111\}
plane.  That is,  experimental
seaweed growths typically involve a significant misorientation from the
\{111\} plane.
The consequences of this will be emphasized below.

\section{RESULTS}

\subsection{Seaweed morphologies}

\label{sec-seaweeds}

Although low anisotropy solidification produces complicated meandering
patterns compared to dendrites, we find noticeable regularity due to
the imposed growth direction and small anisotropies.

\begin{figure}[bt]
\includegraphics[width=2.3in]{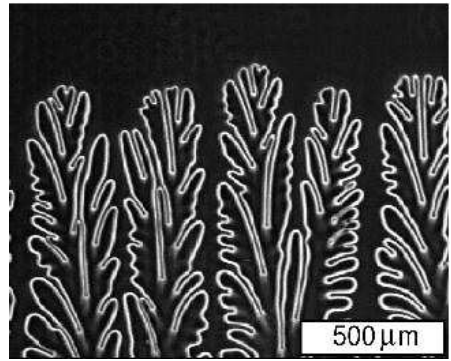}
\caption{
Seaweed growth in 0.25\% SCN-PEO at V = 6.74 $\mu$m/s.  The growth is
composed of seaweed cells, five of which are seen here.
}
\label{seaweed-cells}
\end{figure}

There does
appear to be a typical spacing between the large seaweed cells
\cite{Akamatsu.ea:98:Anisotropy-driven},
as seen in Fig.~\ref{seaweed-cells}.
This spacing is
comparable to that for dendrites grown at the same
conditions (e.g. as in Fig.~\ref{seaweed}),
but is unstable and continuously changes over time.
There is frequent tip splitting and competition between lobes
which are occasionally created or fall behind.  The splitting
events also occur at different places on the tip and create
arms of varying lengths.  These factors lead to
the characteristic meandering and random appearance of the seaweed.

\begin{figure}[bt]
\includegraphics[width=3.3in]{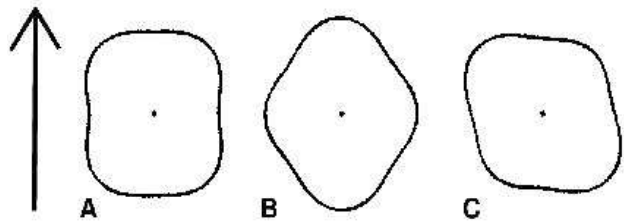}
\caption{
Anisotropic part of surface tension in planes oriented $5^\circ$ from
the \{111\} orientation using the same parameters as in
Fig.~\ref{stens100-111}.
(A) Close to the \{655\} plane
and (B) near \{665\} orientation.  Note that specifying the
plane does not select the orientation with respect to the growth
direction (given by the arrow).  (C) is a specific orientation found by
rotating (B) in the sample plane.
}
\label{stens-off111}
\end{figure}

Given that it is unlikely to randomly seed a seaweed grain within
$1^\circ$ of the \{111\} plane as mentioned above,
we must ask how growth is affected by
small misorientations from the \{111\} plane.
Fig.~\ref{stens-off111} shows a few possible surface
tension profiles for grains misoriented $5^\circ$ from the \{111\} plane
towards the \{100\} or \{110\} orientation and with in-plane rotations.
Not only is the surface tension anisotropic, it is also not generally
fourfold symmetric as usually assumed in simulations and theory.
Since the grains in Fig.~\ref{stens-off111}
are close to \{111\}, the growth will be \textit{relatively}
isotropic and should form seaweeds. However, the dynamics of the seaweed
will depend on the slight anisotropy.

This effect may be particularly
relevant because a slight anisotropy on a nominally
isotropic case will break the symmetry and induce a sense of orientation.
Slight misorientations
from a strongly anisotropic case like the \{100\} orientation will induce
only small changes on the existing profile and will not
be significant.

In particular, if we force a
crystal oriented as Fig.~\ref{stens-off111}A to grow upwards, there
will be a small degeneracy \cite{degeneracy}.
Growth towards the surface tension maxima is preferred and a tip will
tend to grow outwards in both directions leading to a
marked increase in radius or flattening of the tip.
We call this the degenerate seaweed
as there is a small amount of degeneracy which is revealed in the
dynamics.   Forcing a crystal oriented as in
Fig.~\ref{stens-off111}B to grow upwards, the seaweed now
grows along a preferred direction
and the tip will be somewhat more stable than the isotropic seaweed.  This is
the stabilized seaweed.  Fig.~\ref{stens-off111}C shows the same crystal
in Fig.~\ref{stens-off111}B rotated within the plane. As we show below, in this case  upward growth can lead to seaweeds tilted beyond $45^\circ$ as a consequence
of the twofold symmetry.

\begin{figure}[bt]
\includegraphics[width=3.3in]{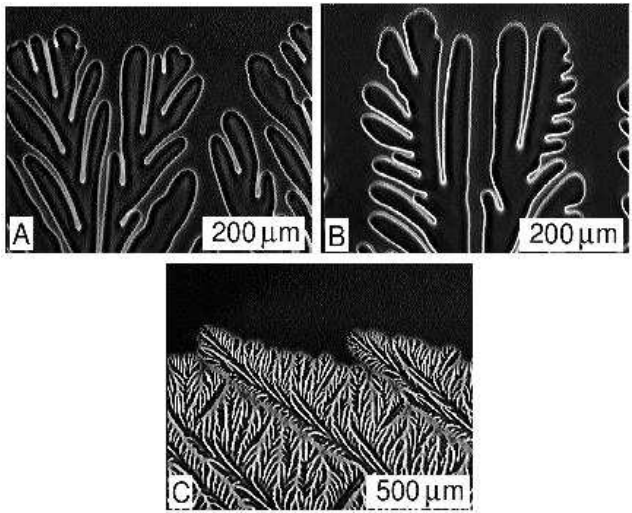}
\caption{
Three kinds of seaweed growths observed in an SCN-ACE sample at
a temperature gradient of 20 K/cm.  (A) A degenerate, or alternating
tip splitting, mode
at $V=8.96 \frac{\mu m}{s}$, (B) a stabilized seaweed at
$V=8.96 \frac{\mu m}{s}$ and (C) a strongly tilted seaweed (tilted beyond
$45^\circ$) at $V=43.6 \frac{\mu m}{s}$ which reveals
a twofold rather than fourfold symmetry.
}
\label{weedtypes}
\end{figure}

Figure \ref{weedtypes} shows a few examples with orientations
similar to those shown in Fig.~\ref{stens-off111}.
In each case, the same sample is used, but each image corresponds to 
grains of different crystalline orientation. 
They are all seaweeds because the tip is unstable to splitting,
but there is a clear qualitative difference in their structure.
We describe these further below.

\subsubsection{Degenerate seaweed, alternating tip splitting}

One of the most striking types of seaweeds is the degenerate seaweed seen in
Fig.~\ref{seaweed}B and Fig.~\ref{weedtypes}A.
At first glance, they appear similar to other
seaweeds, except that the tip is observed to alternately split on the left
and right sides \cite{Utter.ea:01:Alternating,AkamatsuRef-fig20}.
That is, when the tip splits, one of the two new lobes will grow forward
as the other falls behind.  If the lobe towards the left survives, when the
tip splits next, there is roughly
an 85\% chance that the lobe on the right will
survive.

We have characterized this state in detail \cite{Utter.ea:01:Alternating},
including a model which
captures the observed scaling behavior;  The tip splitting
frequency $f$, the wavelength of the tip instability $\lambda_t$,
and the pulling speed $V$ are related as: $f \propto V^{3/2}$,
$\lambda_t \propto V^{-1/2}$, and $f \propto \lambda_t / V$.
The observed frequency exponent of $3/2$ is
identical to what is expected for the  sidebranching frequency in
dendrites\cite{billia-93,Georgelin.ea:98:Onset}.
The relevant results will be summarized
briefly to contrast with other types of seaweeds and to correlate
the previous observations with the surface tension plots shown above:
(a) tip splitting
can regularly alternate, (b) the instability wavelength of tip splitting
is linearly related to the instability wavelength of the planar interface,
and (c) alternating tip splitting is correlated with a strong flattening
of the tip and a particular crystallographic orientation.

\begin{figure}[tb]
\includegraphics[width=2.8in]{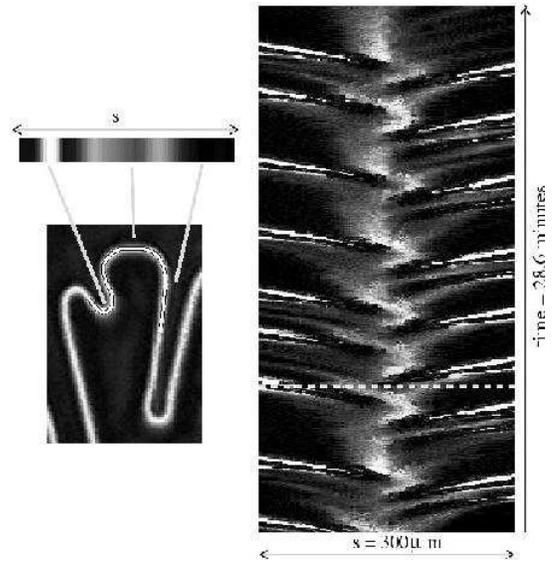}
\caption{
Curvature-time plot for 0.25\% SCN-PEO.
A representation of the curvatures
along the interface near the tip.
To the left is a tip region with an segment indicated in white.  Above it,
the absolute curvatures along this segment are plotted in greyscale.
Stacking sequences of these lines in time
for subsequent pictures gives the curvature-time plot on
the right, where the center line always corresponds to the tip.
Time increases upwards (total time 28.6 minutes).
The width is 300 $\mu$m and the
growth velocity is 2.03 $\mu$m/s.  White corresponds to high curvatures
(radius of curvature less than $\approx 10 \mu$m) and black to zero curvature.
The dashed line indicates the position of the tip that is shown.}
\label{crvprf}
\end{figure}

To gain additional insight, the curvature is measured at each point
on an arc centered on the tip.  Plotting curvature versus the position
along the arc and stacking the plots from successive times, we created
curvature time (CT) plots, as shown in Fig.~\ref{crvprf}.
The arclength \textbf{s} is centered on the tip which is defined as the
furthest point along the growth direction.  The greyscale intensity
corresponds to the absolute value of the curvature.
This plot shows the evolution of the
curvatures in the region of the tip over time.  The center
of the plot always corresponds to the tip.  Each splitting event
is represented by a double line because a deep groove and an additional
tip are created, both of which have high curvature and convect
down the side of the seaweed.

The alternating tip splitting can be very regular as seen in
Fig.~\ref{crvprf}.
There, it is clear that the curvatures at the tip oscillate, reflecting
the alternating flattening and splitting of the tip.
We emphasize that the
periodicity seen in Fig.~\ref{crvprf} is a reflection of the changing
shape of the tip and not an artifact of the tip moving from side to
side since the tip position changes by a relatively small amount.
This is striking because in a relatively isotropic system with a noise
dominated instability such as tip splitting, one expects to find
random and upredictable behavior.
Although there could be a nonlinear feedback
mechanism that leads to an instability such as vortex shedding in fluid
flows, simulations of isotropic solidification
have not revealed such a cycle.
Although rare, this state is not unique, as we have observed it in
three different
samples and it appears to be the quasi-periodicity pointed out in
Fig.~20 of \cite{Akamatsu.ea:95:Symmetry-broken}.

Measurements of the tip instability wavelength $\lambda_t$ versus
the instability of the initially flat front $\lambda_f$ demonstrated an
approximately linear relationship.  This indicates that to a first
approximation, the instability wavelength of tip splitting
arises from the more familiar instability of the flat interface.  The precise
relationships for two particular degenerate seaweed grains show
that $\lambda_t$ is
in fact smaller than $\lambda_f$ ($\lambda_t \approx 0.8 \lambda_f$)
\cite{Utter.ea:01:Alternating}.
The tip will become unstable at
the smallest instability wavelength, since the tip is initially at a
size that is too small to support an instability and grows.  That is, $\lambda_t$ is essentially
probing the small
wavelength branch of this dynamic stability curve.
The evolving tip is more complicated than the initial planar instability
which is itself more complicated than the steady state linear theory
of Mullins and Sekerka \cite{Mullins.ea:64:Stability}.  Despite this,
we find within experimental errors that all of these lengths scale
in the same way as $\lambda \propto V^{-0.5}$.

\begin{figure}[tb]
\includegraphics[width=3.375in]{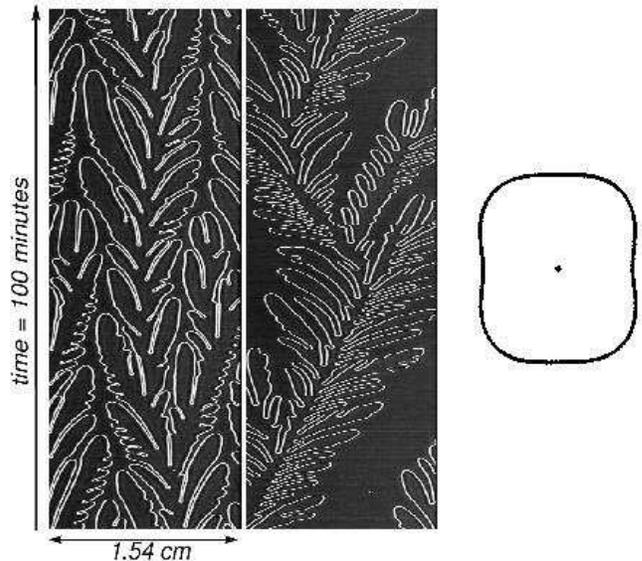}
\caption{
Space-time plot for 0.25\% SCN-PEO degenerate sample.
Time increases upwards.
The growth velocity is 2.71 $\mu$m/s and temperature gradient is
(A) 18 K/cm and (B) 3 K/cm.  The approximate orientation of the grain
is represented by the surface tension plot on the right.
This is the same grain as Fig.~\ref{seaweed}B.
}
\label{tipdegen-st}
\end{figure}

The observed flattening
of the tip is precisely what we might expect if
the crystal was oriented as Fig.~\ref{stens-off111}A.  To verify
that this is the case, we performed a run at a very small temperature
gradient so that
the growth would be dominated by any crystalline anisotropy rather
than the imposed temperature gradient.
With a reduced temperature gradient,
the resulting growth is closer
to that of free growth.
Figure~\ref{tipdegen-st} shows a space-time (ST) plot from the
run (see~\cite{Akamatsu.ea:95:Symmetry-broken}, for example).  
It was created by taking the pixels from
a fixed distance behind the interface
from each image and stacking them sequentially in time
(similar to the CT plot).
The distance behind the interface in this figure is
$\approx 12 \lambda_f$.
The plot is essentially a chart
recording of the growth in the absence of further coarsening.  It's
clear that the growth locks into two particular directions, consistent with
the surface tension profile shown on the right.

A state qualitatively similar to this alternating tip splitting
is observed in viscous fingering experiments,
but is due to an additional perturbation, such as the presence of a
bubble trapped at the tip \cite{Couder.ea:86:Narrow}.
Park and Homsy
also see a near periodic splitting, although
there is not a long enough sequence to be sure \cite{Park.ea:85:instability}.
Alternating tip splitting can also be observed in simulations when
competing anisotropies balance \cite[see Fig.~3c]{Ihle:00:Competition}.  This
might result from a slight degeneracy in a relatively isotropic surface
tension profile as we believe these results show.

At low speeds, the seaweed cells become more stable and lead to a deviation
from the observed $f \propto V^{1.5}$ scaling.
Also, the slight asymmetry in the anisotropy is revealed
and splitting events to one side dominated the splits to the other.
At higher speeds, the structures become smaller and growth is more three
dimensional, making it difficult to extract the interface and follow the tip.

\subsubsection{Stabilized seaweed}

Fig.~\ref{weedtypes}B shows the stabilized seaweed.  Note that it is
the same sample as the degenerate seaweed in Fig.~\ref{weedtypes}A
growing at identical conditions except for the orientation of the
crystal.  Unlike the degenerate seaweed,
the tip is not generally splitting towards alternate sides.  In fact,
the horizontal branches (for example, on the rightmost tip) are true
sidebranches which develop below the tip,
and the tip splitting is much less frequent.

\begin{figure}[tb]
\includegraphics[width=3.3in]{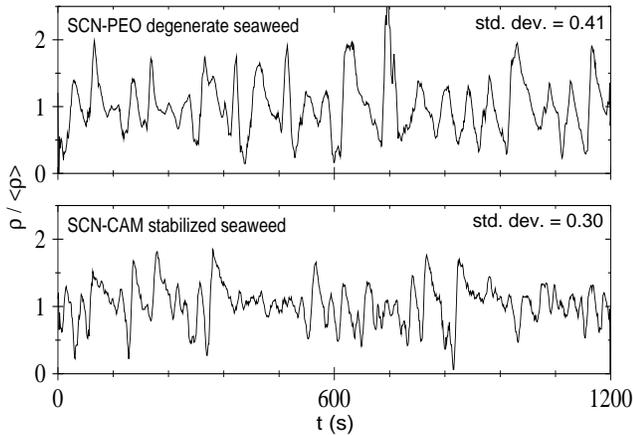}
\caption{
Curvature of the tip $\rho$ for a (A) degenerate seaweed
($V = 2.71 \mu m /s, G = 18 K/cm$) and a (B)
stabilized seaweed ($V = 4.5 \mu m /s, G = 18 K/cm$).
In each case, the curvatures are divided by
the mean for the run $\langle\rho\rangle$.
The standard deviation for the stabilized
seaweed (0.30) is smaller than that for the degenerate seaweed (0.41)
reflecting the increased stability of the tip.
}
\label{tip-curvatures}
\end{figure}

In Fig.~\ref{tip-curvatures},
tip curvature is plotted versus time for typical examples of the
degenerate and stabilized seaweed.  The radius of curvature of the
tip is determined as a function of time where the tip is again defined
as the furthest point along the growth direction.  The curvatures in
each case are normalized by the average for the run.
It's clear that the standard deviation  is smaller for the
stabilized seaweed which confirms that the tip exhibits less variation
in curvature,
suggesting that
this might be an example of the situation shown  in
Fig.~\ref{stens-off111}B.
In contrast,
the degenerate seaweed displays prominent oscillations in curvature
reflecting the continual splitting and flattening of the tip.

\begin{figure}[tb]
\includegraphics[width=3.375in]{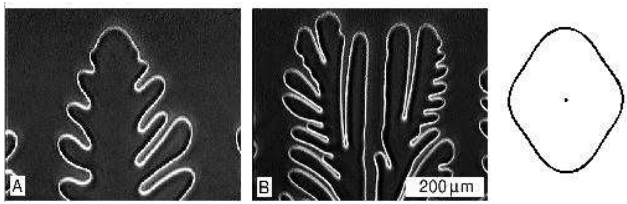}
\caption{
Tip stabilized seaweed at (A) $V = 4.5 \mu m/s$ and (B) $V = 8.96 \mu m/s$
Both images are 1.5\% CN-ACE with G = 20 K/cm.
}
\label{tipstable}
\end{figure}

Using a lower pulling speed, the unstable seaweed growth undergoes a
transition to dendrites, shown in Fig.~\ref{tipstable}.
The resulting growth seen in
Figure~\ref{tipstable}A shows one of an array of dendrites with stable tips,
indicating an anisotropy along the growth direction consistent with the
stabilized seaweed.
Note that this is not simply an artifact of the temperature gradient
constraining the growth, although that might contribute to the stability
of the dendrites.  At corresponding low
velocities, the degenerate state described above appears cellular
but remains  unstable to splitting.

This effect does not appear to be caused by kinetic anisotropy which
generally refers to an increase in anisotropy with increasing velocity.
In fact, this is the opposite effect.  Qualitatively, this
could be interpreted as the same behavior found in
simulations in which competing anisotropies
balance \cite[see Fig.~2 in which decreasing undercooling (2b to 2a) leads
to more ordered growth]{goldenfeld-trans} \cite{Ben-Jacob.ea:87:Interfacial},
but we do not believe anisotropies
in different directions exist in the present experiment.  We also observe this
grain to appear seaweed-like up to $V = 86 \mu$m/s so there does not appear
to be another anisotropic state that dominates at large growth speed.
With this in mind
and given the evidence in Fig.~\ref{tip-curvatures},
we conclude that there is a
small anisotropy along the growth direction.

It is interesting to note that fractal dendrites described by
Brener \ea~appear very similar \cite[see Fig.~5]{Brener.ea:94:Fluctuation}.
In fractal dendrites, although a central stem of the dendrite is still
definable, large noise or low anisotropy leads to occasional tip splitting.

\subsubsection{Strongly tilted seaweeds}

The degenerate and stabilized seaweeds are, in a sense, the two extremes
of what surface tension profiles
will be seen when misoriented from the \{111\}
plane.  Other growths will be combinations of these behaviors with the
additional freedom to rotate the sample in the plane.

Now considering Fig.~\ref{stens-off111}C,
the surface tension is not fourfold symmetric.
In other words, the model of surface tension based on Eqs.~\ref{sstiff-formula}
and~\ref{stens-formula} used most often
in simulations and theories,
$\gamma (\alpha) = \gamma_0 [1  + \epsilon_0 \cos (4 \alpha)]$,
is not valid here.
The lack of complete fourfold symmetry has been noted before
\cite{Muschol.ea:92:Surface,Dougherty:91:Surface}
but is not typically important for dendrite growth. One
consequence is that we can see dendrites growing at angles larger than
$45^\circ$ with respect to the pulling direction, which does not happen
under the assumption of fourfold symmetry.
If the anisotropy is fourfold symmetric,
a dendrite growing at $\alpha > 45^\circ$ will have sidebranches
at $90^\circ - \alpha < 45^\circ$ in the other direction
which will be favored.

Fig.~\ref{weedtypes}C
is an example of this in which a tip splitting growth is tilted at
approximately $53^\circ$, consistent with a surface tension anisotropy
oriented like Fig.~\ref{stens-off111}C.
This picture shows that twofold symmetry can be important in
seaweed growth.
A similar observation can be seen in dendrites
\cite[{see Fig. 25}]{Akamatsu.ea:95:Symmetry-broken} although no
mention is made of the implications of the large tilt angle.

\begin{figure}[tb]
\includegraphics[width=3.3in]{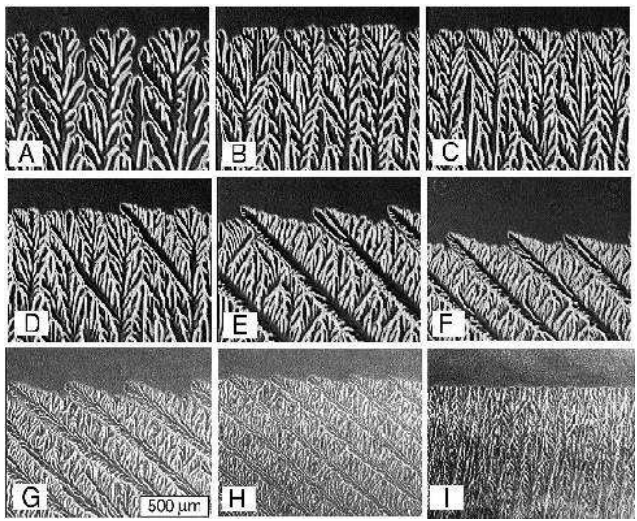}
\caption{
Transition to strongly tilted growth with increasing growth speed.  Images
are shown at pulling speeds of (A) 4.5, (B) 9.0, (C) 13.4, (D) 17.8,  (E) 22.1,
(F) 43.6, (G) 86.4, (H) 182, and (I) 242 $\mu$m/s.  
The sample is 1.5\% SCN-ACE and G = 40 K/cm.
Images H and I show a transition from strongly tilted seaweed back to growth 
oriented along the pulling velocity.  Although reproducible, the large scale linearity of 
the temperature gradient is not maintained at $V > 100 \mu$m/s.
}
\label{stilt-transition}
\end{figure}

In Fig.~\ref{stilt-transition} we show the progression of this strongly
tilted seaweed with increasing growth speed.  At low speeds there is a slight
tilt to the right.  As the pulling speed is increased, branches to
the left are more apparent
until at large enough speeds they dominate the growth.   At much
larger speeds, the seaweed actually
reverts to a slight tilt to the right as seen
at low speeds.   Although this transition was reproducible, the
temperature gradient is far from linear at those speeds and we are not able
to draw reliable conclusions from these observations.

The transition is qualitatively different from the cell to
dendrite transition in which cells gradually tilt further towards the
crystalline axis until the transition to dendrites
\cite{Akamatsu.ea:97:Similarity}.
In that case, the cells smoothly tilt further towards the crystalline
axis, while here the tilted arms grow out from the seaweed with a lifetime
that increases with pulling speed until they become stable.

\subsubsection{Degenerate-stabilized seaweed transitions}

\begin{figure}[tb]
\includegraphics[width=3.375in]{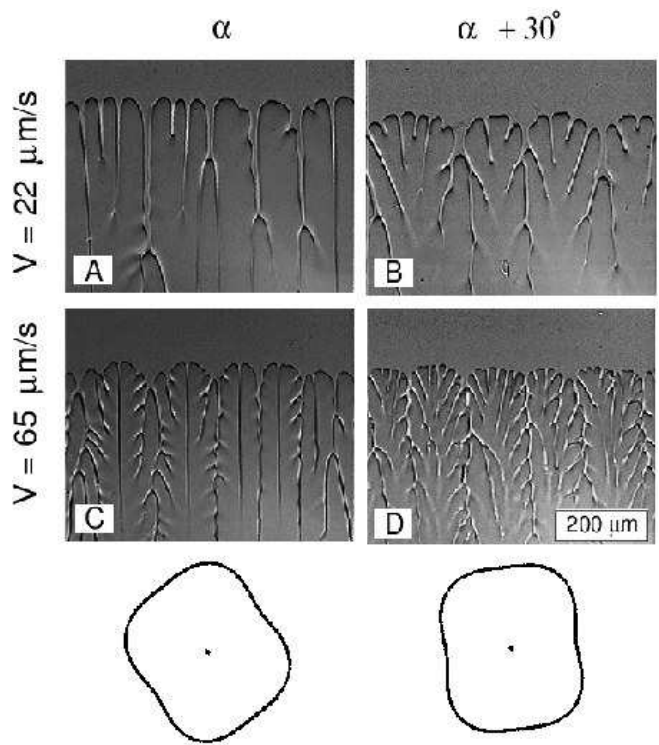}
\caption{
Transition between stabilized and degenerate seaweed growth with
in-plane rotation of sample.  The sample is 0.5\% SCN-PEO.  At a certain
sample orientation ($\alpha$), with (A) V = 22 $\mu$m/s, the sample
grows as stabilized seaweed and at (C) V = 65 $\mu$m/s doublons form.
After rotating the sample by 30$^\circ$, the growth becomes (B)
a degenerate seaweed at V = 22 $\mu$m/s and (D) remains seaweed at
V = 65 $\mu$m/s.  Below, possible surface tension profiles are shown
which are rotated by 30$^\circ$ with respect to each other.
}
\label{rotate-weed}
\end{figure}

From the above discussion, it should be possible to observe transitions
between different seaweed types with an in-plane rotation of the sample.
Fig.~\ref{rotate-weed} shows an example.  At low speeds, a stabilized
seaweed forms (\ref{rotate-weed}A).  When rotated by 30$^\circ$, the
growth becomes a degenerate seaweed and exhibits alternating tip splitting
(\ref{rotate-weed}B).
At higher growth speeds for the same two orientations, stable doublons
(\ref{rotate-weed}C)
become unstable to tip splitting (\ref{rotate-weed}D)
with the same sample rotation.
At the bottom of  Fig.~\ref{rotate-weed}, possible surface tension
profiles are shown which are rotated by 30$^\circ$ with respect to each
other. Doublon growth will be addressed in a future publication 
\cite{doublon-paper}.

\subsection{Fractal dimension}

\label{sec-fractal}

\begin{figure}[tb]
\includegraphics[width=3.0in]{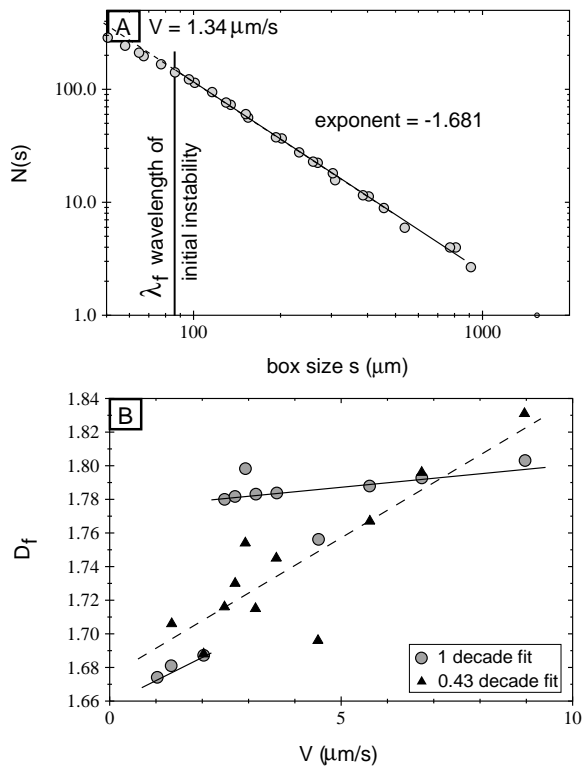}
\caption{Fractal analysis for degenerate seaweed (0.25\% SCN-PEO).
A box counting method is used in which a grid
of spacing \textit{s} pixels
is superimposed on a picture of a dendrite and the number of
boxes containing any part of the interface (N(\textit{s})) is counted.
(A) A linear
region on a log-log plot indicates a fractal range with the dimension given
by the magnitude of the 
slope.  The plot here is for a growth velocity of 1.34 $\mu$m/s and the
experimental initial instability wavelength is included as the lower length
scale cutoff for the fractal range.
(B)  Averaging results from 1000 pictures for each point, the
fractal dimension versus the pulling speed is plotted.
The solid line (circles, 1 decade fit) suggests a discontinuous jump while the
dashed line (triangles, 0.43 decade fit) suggests a smooth transition.}
\label{fractal}
\end{figure}

Since we expect to see a crossover from fractal to compact structures with
increased pulling speeds \cite{brener-96-8}, we measured the
fractal dimension $D_f$ of our
images by using a standard box counting method
described in Fig.~\ref{fractal}
\cite{peitgen92}.  The lower physical cutoff of the fractal range is
close to the wavelength of the initial
instability of the flat interface $\lambda_f$.
The experimental measurement of this
value has been measured at each pulling speed and is
indicated on the plot (Fig.~\ref{fractal}A).
$D_f$ is measured as the magnitude of the
slope for box sizes s $> \lambda_f$.

Figure~\ref{fractal}B shows the fractal dimension versus the pulling speed
for a degenerate seaweed.
The circles correspond to fitting over one decade on Fig.~\ref{fractal}A
to determine $D_f$.
The triangles correspond to fitting over 0.43 decades, equivalent
to one division on a natural log plot, which has been used in some
previous results \cite{ihle-mk-93-4}.
It's clear that the fractal dimension is
sensitive to the range of data taken for the fit, although the general
trend seems to remain that the fractal dimension increases with
pulling speed.  This increase from close to the diffusion limited
aggregation value of 1.67 towards 2 would be consistent with Brener \ea's
prediction of a noisy transition from fractal to compact growth. In addition, 
Brener~\ea predicted that the transition is discontinuous. 
When using data from a fit over one decade we observe such a discontinuous 
jump, however, fitting over a shorter region does not show such a jump. 
The fit is taken starting at $\lambda_f$, i.e. the fit over one decade 
includes box lengths between $\lambda_f$ and 10$\lambda_f$.

At first, Figure~\ref{fractal}B  looks promising
in indicating a transition from fractal to compact growth, but a
few important issues must be noted.
As mentioned, the slope is sensitive to the range of the
fit and, at most, one decade in length scales can be used.
In other words, these
pictures do not exhibit growth that is clearly fractal over a significant
range of length scales.
We question whether previous experiments have had the same limitations.
At lower speeds, as the seaweed tends towards less developed cellular
growth, the calculated dimension actually drops
towards one rather than levelling out.
The fractal dimension also appears to be most well defined at the
tip, as the dimension increases towards 2 when more of the deep groove
region is included in the analysis.
This could be an artifact of the imposed gradient and
may not be an issue in free radial growth where the
number of lobes must continually increase.

In summary, our results suggest a transition from fractal to compact
growth, but we find that the range of data spans only one decade, making a conclusive interpretation as  fractal scaling impossible.

\subsection{Seaweed-dendrite transitions}

\label{sec-transitions}

\begin{figure}[tb]
\includegraphics[width=3.0in]{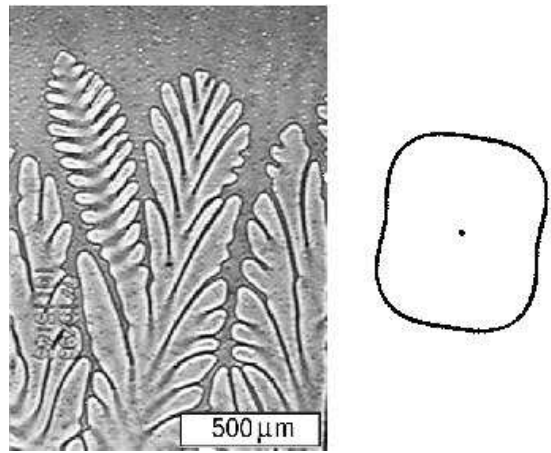}
\caption{
Dendritic growth from a degenerate seaweed.  The approximate
orientation of the crystal is inferred to be that represented by the
surface tension plot on the right.  The sample is 0.25\% SCN-PEO growing at
V = 4.5 $\mu$m/s and G = 30 K/cm.
}
\label{dendfromwd}
\end{figure}

In low anisotropy growth, it is possible to observe dendritic growths in
patterns that otherwise are seaweed. For example, 
Figure~\ref{dendfromwd} shows a snapshot of the alternating tip splitting
seaweed that is tilted to the right approximately 9 degrees as represented
by the surface tension plot.  
One of the side branches  of the seaweed has
nucleated  a dendritic branch. 
Assuming that the anisotropy of the
crystal is mirror symmetric, the angle between the dendritic branches would be
$43^\circ$, which is consistent with the value of $40^\circ$
for similar branches in Fig.~\ref{tipdegen-st}B.  
Due to the
regularity of the sidebranches these dendrites  look  different from the 
usual  dendrites which are observed for growth along the crystal's easy 
axis.   They look very
similar to the tip oscillating growth or symmetric tip splitting state
of Honjo \ea \cite{Honjo.ea:85:New}.  In their results, the tip velocity
and curvature oscillate in time, but these oscillations are not apparent here.

\begin{figure}[tb]
\includegraphics[width=3.0in]{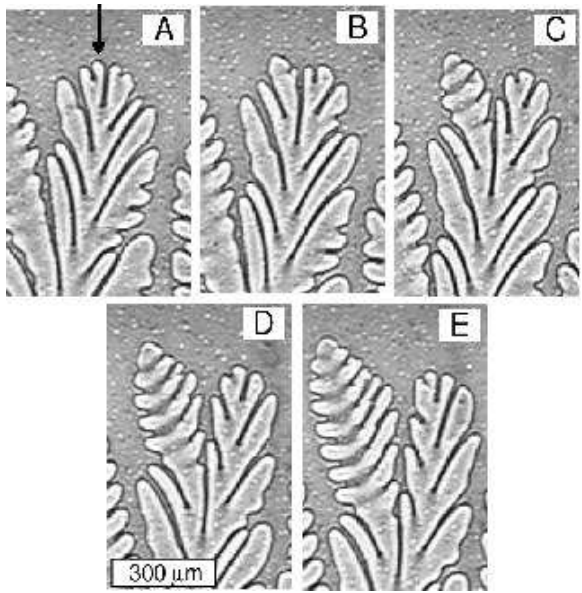}
\caption{
The formation of a dendritic growth like that shown in Fig.~\ref{dendfromwd}.
The arrow indicates the seaweed arm that develops into the dendritic branch.
The time between pictures is 30 s.
The sample is 0.25\% SCN-PEO growing at
V = 4.5 $\mu$m/s and G = 30 K/cm.
}
\label{wdoutgrowth}
\end{figure}

\begin{figure}[tb]
\includegraphics[width=3.0in]{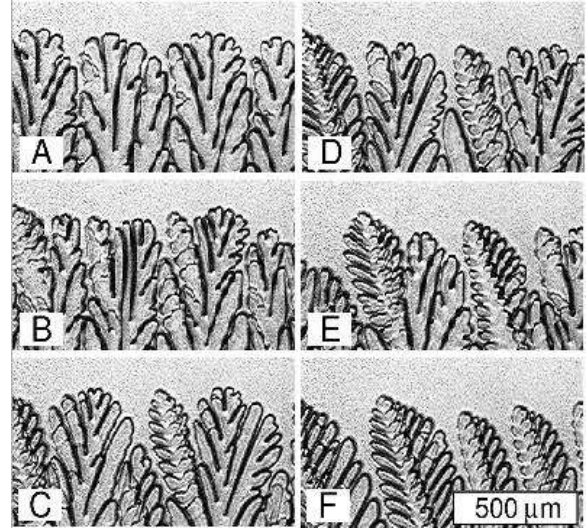}
\caption{
Transition from a (A) seaweeed morphology to a (F) dendritic morphology
over time.  The pictures are separated by 240 seconds.
The sample is 0.25\% SCN-PEO growing at
V = 4.5 $\mu$m/s and G = 45 K/cm.
}
\label{w2dfront}
\end{figure}

Fig.~\ref{wdoutgrowth} shows the time evolution of the formation of one of these dendritic
branches.  The arrow highlights the seaweed arm which becomes dendritic.
We also observe in Fig.~\ref{dendfromwd}A and Fig.~\ref{wdoutgrowth}E
that the dendritic branch grows ahead of the
seaweed growth.  This is not surprising as it is already known that
dendrites grow faster than seaweeds at the same conditions.
One might guess that the dendritic arm could
grow ahead of the neighboring seaweed and dominate the growth.   Indeed,
the seaweed growth in this case is not stable --  dendritic
branches  nucleate at different points along the interface and take over
the pattern.  Both seaweeds and dendrites can be understood as two stable states
 with dendrites being dynamically preferred over seaweeds.
The  seaweeds are typically found to be stable until the first dendrites are formed.
 An example of the the evolution of the seaweed to dendrite transition 
 is shown in Fig.~\ref{w2dfront}.  
There, an initial seaweed is seen to nucleate dendritic branches. 
In Fig.~\ref{w2dfront}E, after about 20 minutes of growth,
some of the dendrites have managed to grow ahead of
the seaweeds .

\begin{figure}[tb]
\includegraphics[width=2.8in]{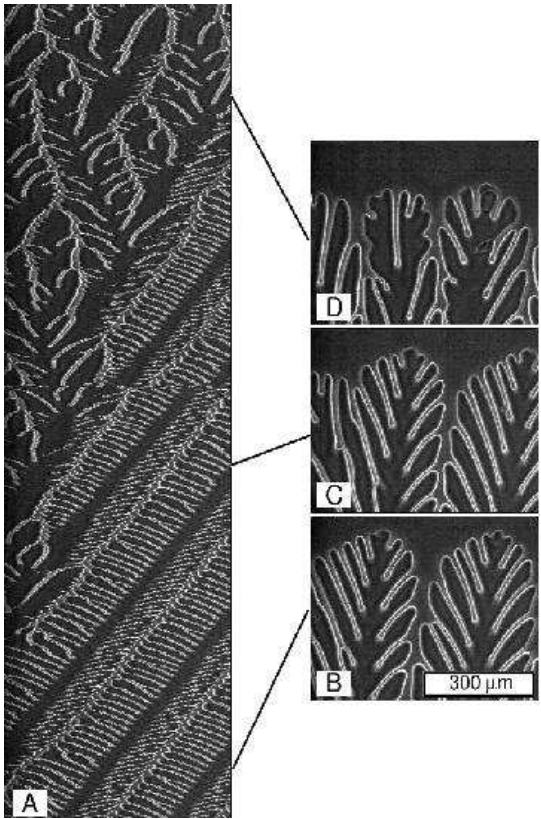}
\caption{
(A) Space-time (ST) plot of a dendritic interface that undergoes a transition
from a (B) dendritic to
a (D) seaweed morphology.  The discontinuity in the ST plot is due to
a translation of the microscope stage along the interface.
The sample is 0.12\% SCN-CAM growing at
V = 13.4 $\mu$m/s and G = 22 K/cm.
}
\label{dend2weed}
\end{figure}

A seaweed cannot generally overtake a dendrite since it grows at larger
undercooling.   It is possible though when the dendrite is angled away
from the seaweed.  Fig.~\ref{dend2weed} shows a space-time plot in which
a dendritic growth appears stable for a long time.  After a number of
failed attempts, a seaweed branch nucleates on the left and gradually
spreads to the right.   It is clear from the ST plot that the seaweed
grows out from a branch on the dendritic growth and is not simply
another grain.
Fig.~\ref{dend2weed}C shows the initial formation of the seaweed grain.

\section{CONCLUSION}

\label{sec-conclusions}

In conclusion, we find that misorientations from the \{111\} plane
lead to different types of seaweeds arising from small anisotropies.
These include the
degenerate, stabilized and strongly tilted seaweeds.  The degenerate and
stabilized seaweeds are the two basic types of misorientation with the
additional freedom to have in-plane rotations.
The strongly tilted
growth in particular highlights the underlying twofold, rather than
fourfold, symmetry.  The degenerate state is found to allow a
regular alternating tip splitting to develop. The observed growth 
morphologies are correlated with the plots of the in-plane surface 
tension. 

The fractal dimension was
studied as a function of growth velocity for the degenerate seaweed.
Although we observe
a general trend supporting the predicted fractal to compact transition
with increasing undercooling, there is not a sufficient scaling region
in directional solidification to consider it to be a true fractal.
Transitions between seaweed and dendrite growth were also observed. 

Ultimately, the question is:
How does surface tension anisotropy select
particular growth morphologies?
In particular, we ask:
(i) What can we learn about the crossover between
tip splitting and sidebranching with small
increasing (non-fourfold) anisotropies?
(ii) How can we elucidate the role and identify the relative importance of
kinetic anisotropy? and 
(iii) Are similar morphologies observable in other low anisotropy systems?

This work was supported by the Cornell Center for Materials Research (CCMR),
a Materials Research Science and Engineering Center of the National Science
Foundation (DMR-0079992).



\begin{thebibliography}{10}
\bibitem{random-seeding}
For a sphere of radius 1, the surface area of a cap composed of all points
  within $1^\circ$ of a particular vector, e.g. the 111 vector, is $A_1 = 2 \Pi
  (1 - \cos (1^\circ)) = 1.52 \times 10^{-4}$. There are 8 such equivalent
  orientations and the total area of all these caps is $A = 8 \times A_1 = 1.22
  \times 10^{-3}$. Taking the probability of randomly selecting one of these
  orientations to be the ratio of this area to the area of the sphere, we get
  $P = \frac{A}{4 \Pi} \approx \frac{1}{1600}$.
\bibitem{degeneracy}
Degeneracy refers to an orientation in which growth along $\pm 45$ degrees
  relative to the pulling velocity is favored, leading to competition between
  the two preferred orientations (e.g. when Fig.~\ref{stens100-111}A is rotated
  by $45^\circ$ and the growth direction is upwards).
\bibitem{AkamatsuRef-fig20}
See reference \cite{Akamatsu.ea:95:Symmetry-broken}. In particular, note Fig.
  20 which highlights a region of ``quasiperiodicity'' similar to these
  results.
\end{thebibliography}


%
%

%
%

\end{document}